\documentstyle[12pt]{article}
\begin{document}

\def\slash{{\rlap /}}
\def\PRL{Phys. Rev. Lett. }
\def\be{\begin{equation}} \def\bea{\begin{eqnarray}}
\def\ee{\end{equation}}\def\eea{\end{eqnarray}}
\def\ncr{\nonumber\\ }
\def\tr{{\rm tr}\,}\def\STr{{\rm STr}\,}
\def\Sdet{{\rm Sdet}\,}\def\Tr{{\rm Tr}\,}
\def\dos{\stackrel{\star}{,}}
\def\DG{\Delta ^{\a\b\g}_{\s\r\m}\,}
\def\DR{\Delta ^{abc}_{srm}\,}
\def\nn{{\Box ^{-1}N_0}} \def\nj{{\Box ^{-1}N_1}}
\def\tj{{\Box ^{-1}T_1}} \def\td{{\Box ^{-1}T_2}}
\def\cpe{{i\over (4\pi )^2\epsilon} }
\def\jcpe{{1\over (4\pi )^2\epsilon} }

\def\MBVR{Maja Buri\'c
\footnote{E-mail: majab@ff.bg.ac.yu} and
 Voja Radovanovi\'c\footnote{E-mail: rvoja@ff.bg.ac.yu}\\
{\it Faculty of Physics, P.O. Box 368, 11001 Belgrade,
Yugoslavia}}

\def\endtitle{\par\end{quotation}\vskip3.5in minus2.3in\newpage}


\def\a{\alpha}
\def\b{\beta}
\def\c{\chi}
\def\d{\delta}
\def\e{\epsilon}                
\def\f{\phi}                    
\def\g{\gamma}
\def\h{\eta}
\def\i{\iota}
\def\j{\psi}
\def\k{\kappa}
\def\l{\lambda}
\def\m{\mu}
\def\n{\nu}
\def\o{\omega}
\def\p{\pi}                     
\def\q{\theta}                  
\def\r{\rho}                    
\def\s{\sigma}                  
\def\t{\tau}
\def\u{\upsilon}
\def\x{\xi}
\def\z{\zeta}
\def\D{\Delta}
\def\F{\Phi}
\def\G{\Gamma}
\def\J{\Psi}
\def\L{\Lambda}
\def\O{\Omega}
\def\P{\Pi}
\def\Q{\Theta}
\def\S{\Sigma}
\def\U{\Upsilon}
\def\X{\Xi}
\def\pa{\partial}
\def\de{\nabla}

\def\ca{{\cal A}}
\def\cb{{\cal B}}
\def\cc{{\cal C}}
\def\cd{{\cal D}}
\def\ce{{\cal E}}
\def\cf{{\cal F}}
\def\cg{{\cal G}}
\def\ch{{\cal H}}
\def\ci{{\cal I}}
\def\cj{{\cal J}}
\def\ck{{\cal K}}
\def\cl{{\cal L}}
\def\cm{{\cal M}}
\def\cn{{\cal N}}
\def\co{{\cal O}}
\def\cp{{\cal P}}
\def\cq{{\cal Q}}
\def\car{{\cal R}}
\def\cs{{\cal S}}
\def\ct{{\cal T}}
\def\cu{{\cal U}}
\def\cv{{\cal V}}
\def\cw{{\cal W}}
\def\cx{{\cal X}}
\def\cy{{\cal Y}}
\def\cz{{\cal Z}}


\def\CMP{Commun. Math. Phys.}
\def\NP{Nucl. Phys. B\,}
\def\PL{Phys. Lett. B\,}
\def\PR{Phys. Rev. Lett.}
\def\PRD{Phys. Rev. D\,}
\def\CQG{Class. Quant. Grav.}
\def\IJMP{Int. J. Mod. Phys.}
\def\MPL{Mod. Phys. Lett.}



\topmargin=.17in                        
\headheight=0in                         
\headsep=0in                    
\textheight=9in                         
\footheight=3ex                         
\footskip=4ex           
\textwidth=6in                          
\hsize=6in                              
\parindent=21pt                         
\parskip=\medskipamount                 
\lineskip=0pt                           
\abovedisplayskip=1em plus.3em minus.5em        
\belowdisplayskip=1em plus.3em minus.5em        
\abovedisplayshortskip=.5em plus.2em minus.4em  
\belowdisplayshortskip=.5em plus.2em minus.4em  
\def\baselinestretch{1.2}       
\thicklines                         
\oddsidemargin=.25in \evensidemargin=.25in      
\marginparwidth=.85in                           


 Paper

\def\title#1#2#3#4{
        {\hbox to\hsize{#4 \hfill  #3}}\par
        \begin{center}\vskip.5in minus.1in {\Large\bf #1}\\[.5in minus.2in]{#2}
        \vskip1.4in minus1.2in {\bf ABSTRACT}\\[.1in]\end{center}
   
     \begin{quotation}\par}
\def\author#1#2{#1\\[.1in]{\it #2}\\[.1in]}
 A4

\def\endabstract{\par\end{quotation}
        \renewcommand{\baselinestretch}{1.2}\small\normalsize}

\title{The one-loop effective action for quantum electrodynamics on noncommutative space }
{\MBVR}{}{July 2002}

\noindent In this paper we calculate the divergent part of the one
loop effective action for QED on noncommutative space using
the background field method. The effective action is
obtained up to the second order in
the noncommutative parameter $\theta$ and in the classical
fields.

\endtitle

\section{Introduction}

The discovery of noncommutative structures in string theory gave a
considerable boost to the noncommutative field theories in the
last few years, and put them into the focus of  vigorous
investigation. Besides the research on the various models
appearing in the context of string theory,  a lot of work has
been devoted to the analysis of noncommutative (NC) theories for
themselves. One approach is to formulate the theory in a
representation-free manner, i. e. to build an abstract algebra of
noncommuting coordinates with a defined set of relations, and then
endow it with structures like derivations, forms, fields etc.
\cite{mssw, jmssw, mad}. The most commonly
discussed is the canonical structure, defined by
 \be\left[ \hat x^\mu , \hat x^\nu \right]=i\theta ^{\m\n}
 \ ,\label{can}\ee
$\hat x^\m$, $\hat x^\n$ are the elements of the algebra and
 $\theta^{\m\n}=-\theta^{\n\m}$ are constant complex numbers.
The other structures like Lie algebra or quantum
plane have been treated as well \cite{mssw, jmssw}. The other
possibility is to consider the
representation of NC theory by the
 fields on commutative space, encoding the noncommutativity  in
 the definition of multiplication. The multiplication which
 corresponds to the canonical structure (\ref{can})
is the so-called Moyal-Weyl or $\star$-product: \be\label{moyal}
f\star g=e^{{i\over 2}\theta^{\m\n}{\pa\over \pa x^\m}{\pa\over
\pa y^\m}}f(x)g(y)|_{y\to x}\ ,\ee
where $f$ and $g$ are functions of the coordinates $x^\m$.
Obviously, \be\label{can1}
[x^\m \dos x^\n]=x^\m\star x^\n-x^\n\star x^\m = i\theta^{\m\n} \
.\ee

The   $\star$-product with its properties in integration
  provides  a new class of
 actions characterized by dimensionfull parameter $\theta$, which
 have nonlocal lagrangians. In this setting the definition of
  noncommutative
 scalar field theories like $\Phi ^3$ or $\Phi ^4$ is
 straightforward. However, if one wants to define
a gauge theory, the use of noncommutative multiplication rule
imposes severe restrictions both on the choice of the gauge group
and on the choice of its representation \cite{h,bsst,cpst}. For
example, in NC electrodynamics  the values of  charge are
quantized and restricted to $\pm 1, 0$.

The result of Seiberg and
Witten \cite{sw} on the equivalence of certain commutative and
noncommutative
gauge theories shows that  noncommutativity  is not
equivalent to  quantization.  Noncommutative field theories can be
quantized  in the conventional perturbative way \cite{kw, mrs, mst}
 and  their properties as unitarity and renormalizability
can be analyzed (see \cite{sz,dn} for a more complete list of
references). Several novel features appear, e.g. UV and IR sectors
are mixed in the perturbative expansion. The UV-IR mixing can  be
seen from the fact that 'nonplanar' diagrams contain terms
proportional to $\vert\tilde p\vert ^{-n}$, with $\tilde p^\mu
=\theta ^{\mu\nu}p_\nu$.

Apparent nonrenormalizability of the NC theory might, in
principle, be regained after the summation of the perturbation
series. This was indeed shown for the NC $\Phi ^4$ theory in
\cite{gp}. However, it is a nontrivial task to prove a similar
assertion for  noncommutative U(1).

An obvious drawback of the perturbative treatment of NC theories
 is that the results, expressed as $\vert\tilde p\vert ^{-n}$,
are also nonperturbative in the  parameter $\theta$. Therefore
one cannot make a smooth commutative limit or estimate the
effects of noncommutativity in the lowest order (in the sectors
where they are small). In order to deal with this problem
one may try
 to use the complementary approach of \cite{mssw}.
Namely, it was shown in \cite{mssw} that, starting with the
general abstract algebra, one can construct the representation of
gauge symmetry for arbitrary group using the concept of 'covariant
coordinates'. The gauge parameter and the gauge potential of the
group are not Lie-algebra-valued any more, but take values in the
enveloping algebra. If the generators of the gauge algebra are
denoted by  $T^a$,
 the following expansion of
the gauge parameter $\hat\Lambda $
in powers of symmetrized products of $T^a$ holds:
 \be
 \hat\Lambda = \Lambda _aT^a + \Lambda ^1_{ab} : T^a T^b: + \dots +
\Lambda ^{n-1}_{a_1
  \dots a_{n}} : T^{a_{1}}\dots T^{a_n} : + \dots \  .
\ee The gauge and matter fields are expanded in a similar way,
\cite{jmssw}. The interesting point is that this expansion
coincides with the expansion in $\theta$, if one maps the NC
fields into the fields on commutative space via the Seiberg-Witten
(SW) map \cite{sw}.
 The coefficients in the expansion depend
on the derivatives of  gauge and matter fields,
so in some sense NC gauge theory is nonlocal both in spatial and
gauge directions.

The SW map induces the $\theta$-expansion of the action as well.
 This gives the possibility to treat the $\theta$-linear
term in the action as the first correction which describes the
effects of noncommutativity in the lowest order.  Further,
$\theta$-expanded action enables one to approach the problem of
quantization in a different way, i.e. considering lagrangian order
by order in $\theta$. In this context,  nonuniqueness of the SW
map \cite{ak}  takes the role of an additional 'symmetry' which
might be used to prove renormalizability. This was done  for
noncommutative U(1) in \cite{bggpsw}. The use of SW map also opens
the possibility to circumvent the restrictions on the values of
charges, e.g. in NC standard model \cite{cjsww}.

Renormalizability of noncommutative  QED with fermions
 seems to be a more complicated problem. It was discussed
extensively in the papers \cite{bggpsw, bgpsw1, bgpsw2} and one of
their conclusions was that it would be of vital importance to find
the exact form of all divergent $\theta$ and $\theta ^2$
counterterms in the action.
 In the recent paper \cite{w}, Wulkenhaar  found
all  $\theta$-linear divergencies in 2-, 3-, and 4-point functions
diagramatically.

In this paper we calculate the divergent part of the one-loop
effective action in the second order in the noncomutative
parameter $\theta$ and in the classical fields, for noncommutative
QED. We use the background field method. The plan of
the paper is the following. In section 2 we briefly define our
notation and fix the form of the lagrangian in the first order in
$\theta$. In section 3 we give the basics of the background field
method and outline how it will be applied to the given model.
Section 4 presents the results for the
effective action, while the appendix contains some helpful
formulas.

\section{Classical theory}

The noncommutative space which we use is  ${\bf R}^4$ with the
canonical structure \be\ [x^\m \dos x^\n]= i\theta^{\m\n} \ ,\ee
where $\m ,\n =0,\dots 3$
 and
$\star$ is the Moyal-Weyl product  (\ref{moyal}). The classical
action for electrodynamics  on this space is given by \be
\label{Snc}S=\int d^4x\hat{\bar \psi}\star (i\g^\m\hat
D_\m-m)\hat\psi -{1\over 4}\int d^4x \hat F_{\m\n}\star\hat
F^{\m\n}\ .\ee
Here, $\hat\psi$ is the noncommutative fermionic
matter field while
 $\hat A_\m $ is the gauge potential. The corresponding
 field strength  $\hat F_{\m\n}$  is defined as \be \hat
F_{\m\n}=\pa_\m\hat A_\n-\pa_\n\hat A_\m-i(\hat A_\m\star\hat
A_\n-\hat A_\n\star\hat A_\m)\ ,\ee
and the covariant derivative $\hat D_\m\psi$ is  \be \hat
D_\m\hat\psi = \pa_\m\hat\psi -i\hat A_\m\star \hat \psi\ .\ee It
is clear that this theory is nonlocal.

The fields $\hat\psi $, $\hat A_\m$, $\hat F_{\m\n}$ which give a
representation of noncommutative electrodynamics can be, via the
SW map,  mapped
 into the representation of ordinary U(1). To
the first order in $\theta$ the map is given by \cite{jmssw}:
  \be\label{sw} \hat A_\m=A_\m-{1\over
2}\theta^{\r\s}A_\r(\pa_\s A_\m+F_{\s\m})\ ,\ee
 \be \label{spinor} \hat\psi =\psi-{1\over
2}\theta^{\m\n}A_\m\pa_\n\psi \ . \ee
  Inserting (\ref{sw}) and (\ref{spinor}) into the action (\ref{Snc}),
  we get  the classical $\theta$-expanded action
 \cite{jmssw, bggpsw}
\be S =S_0
+S _{1,A}+S_{1,\psi}\ ,\ee with
 \be\label{l0} S _0 =\int
d^4x\,\Big[\bar\psi \big( i\g ^\m D_\m-m\big)\psi -{1\over
4}F^{\m\n}F_{\m\n}\Big]\ ,\ee
\be \label{l1A}S _{1,A}=-{1\over
2}\,\theta ^{\r\s}\int d^4 x\,\big[ F_{\m\r}F_{\n\s}F^{\m\n}-{1\over
4}F_{\r\s}F_{\m\n}F^{\m\n}\big]\ ,\ee
\be
\label{l1psi}S_{1,\psi}={1\over 2}\,\theta ^{\r\s}\int d^4x\,\Big[ -i
F_{\m\r}\bar\psi\g ^\m D_\s\psi +{1\over 2}F_{\r\s}\bar\psi (-i\g
^\m D_\m+m)\psi \Big] \ .\ee The 'commutative' covariant
derivative is $D_\m \psi =\pa _\m\psi-iA_\m\psi\ $.

For the purpose of functional integration in the next section, we
will  express the Dirac spinor in terms of the Majorana spinors.
They are  introduced as   $\psi _{1,2}={1\over 2}(\psi\pm\psi
^C)$, where $\psi ^C=C\bar\psi^T$ is the charge-conjugated spinor.
The Dirac spinor is  $\psi = \psi _1 +i\psi _2\ $
 \footnote{There is a number of
identities which Majorana spinors satisfy and which we use. To
encounter some of them: let $\phi$, $\chi$ be Majorana spinors.
Then: $\bar\phi\chi =\bar\chi\phi$ ; $\bar\phi\g _\m\chi
=-\bar\chi\g _\m\phi$; $\bar\phi\s _{\m\n}\chi
=-\bar\chi\s_{\m\n}\phi$; $\bar\phi\g _5\chi =\bar\chi\g _5\phi$;
$\bar\phi\g _\m\g _5\chi =\bar\chi\g _\m\g _5 \phi$. } .

The action in terms of Majorana spinors reads:

\bea S _0 &=& \int d^4x\Big[\bar\psi _1 \big( i\g ^\m \pa _ \m -m
\big)\psi _1 + \bar\psi _2\big( i\g ^\m \pa _\m - m \big)\psi
_2\ncr &+& i\bar\psi _1\g ^\m A_\m\psi _2-i\bar\psi _2\g ^\m
A_\m\psi _1-{1\over 4} F^{\m\n}F_{\m\n}\Big] \ ,\eea

\bea  S _{1,\psi } &=& {1\over 2}\, \theta ^{\r\s} \int d^4x
\Big[\big( -{i\over 2}\bar \psi _1\g ^\m \big( F_{\m\r}\pa _\s +
F_{\r\s}\pa _\m + F_{\s\m}\pa _\r \big) \psi _1 + {1\over 2}\, m
F_{\r\s}\bar \psi _1 \psi _1 \nonumber\\ &-& {i\over 2}\, \bar
\psi _2\g ^\m \big( F_{\m\r} \pa _\s + F_{\r\s} \pa _\m + F_{\s\m}
\pa _\r \big) \psi _2 + {1\over 2}\, m F_{\r\s}\bar \psi _2 \psi
_2 \ncr &-& {i\over 2}\bar \psi _1\g ^\m \big( F_{\m\r} A_\s +
F_{\r\s}A_\m + F_{\s\m} A_\r \big) \psi _2  \\ \nonumber
&+&{i\over 2}\bar \psi _2\g ^\m \big( F_{\m\r} A_\s + F_{\r\s}A_\m
+ F_{\s\m} A_\r \big) \psi _1 \Big] \ .\nonumber \label{llp}\eea
The cyclic combinations which appear in
(\ref{llp}) will be  in the following written in a compact way
$$ F_{\m\r}A_\s + F_{\r\s}A_\m +F_{\s\m}A_\r ={1\over 2}\,\Delta
^{\a\b\g}_{\s\r\m}F_{\g\b}A_\a \ ,$$
$$  F_{\m\r}\pa _\s +
F_{\r\s}\pa _\m +F_{\s\m}\pa _\r ={1\over 2}\,\Delta
^{\a\b\g}_{\s\r\m}F_{\g\b}\pa _\a \ ,$$  introducing the symbol
$\DG$. $\DG$ is cyclic separately in upper and lower indices, and
antisymmetric in any pair of upper or lower indices: \be \Delta
^{\a\b\g}_{\s\r\m} =\d ^\a_\s\d^\b_\r\d^\g_\m -\d
^\a_\r\d^\b_\s\d^\g_\m +({\rm cyclic\ }\a \b \g  )=-\epsilon
^{\a\b\g\l}\epsilon _{\s\r\m\l}\ .\ee

\section{Background field method}

In order to find the divergent part of the one-loop effective
action we will use the background field method. Let us introduce
it briefly \cite{mr4,mr3,bv}. If we consider a theory described by
a set of fields $\phi ^i$ with the classical action $S[\phi ^i] $
and sources $J_i$, the generating functional $W[J_i]$ for the
connected Green functions is given by \be \label{W}
e^{iW[J_i]}=\int \prod \cd\phi _ie^{i\Big(S[\phi ^i] +\int dx
J_i\phi^i \Big)}\quad . \ee  The field components $\phi_i$ can be
fermionic or/and bosonic.
 The effective action, \be \label{lt} \G [\phi ^i_{0}]=W[J_i]-\int
dx J_i\phi ^i_0 \ ,\ee is a Legendre transformation of
the generating functional $W[J_i]$. $\phi ^i_0$ are the solutions
of the classical equations of motion: \be J_i=-{\d_R \G\over
\d\phi^i_0}=-{\d_R S\over \d\phi^i_0}\ ,\ee where $\d_R/\d$ is
the right functional derivative.  From (\ref{W}) and (\ref{lt}) we get
 \be
\label{G} e^{i\G[\phi ^i_0]}=\int\prod \cd\phi _i e^{i\Big(S[\phi
^i] +\int dx J_i(\phi^i-\phi ^i_0)\Big)}\quad. \ee In order to
evaluate this integral we decompose the field  $\phi^i$ as
\be\phi^i=\phi ^i_0+\Phi^i \ , \ee where $\Phi_i$ are the quantum
fluctuations around the classical configuration. The functional
integral (\ref{G}) is then calculated by the saddle-point method
after the Taylor expansion:
 \be\label{eG}
e^{i\G [\phi^i_0]}\approx e^{iS[\phi^i_0]} \int \prod \cd\Phi
_ie^{{i\over 2} \int\Phi^iS^{(2)}_{ij}(\phi _0) \Phi^j}\ , \ee
where $S^{(2)}_{ij}={\d_L \over \d \phi^i}{\d_R\over
\d\phi^j}S\Big|_{\phi ^i=\phi^i_0}.$ The result is the one-loop
effective action \be \label{G1} \G [\phi^i_0] =S[\phi^i _0]-{1
\over 2i} {\rm STr} \big(\log S^{(2) }[\phi ^i_0]\big)\ , \ee
\noindent ${\rm STr}$ is the functional supertrace.

 Let us discuss how to apply the background
field method in the case of NC electrodynamics. The fields in the
theory are the real vector  $A_\m$ and the Dirac spinor $\psi$ and
they are coupled. In order to perform the functional integration
we have to put them into one 'multiplet'  field.
However,  $A_\m$ are real-number valued
while $\psi$ are complex-Grassmann valued (if they were
independent they
 would have entered  the effective action with different
coefficients $-{1\over 2}$ and 1).
To make all fields 'real' we need to
express Dirac spinor $\psi$ in terms of two Majorana spinors
$\psi _1$ and $\psi _2$.

Denoting now the quantum corrections now by $\ca ^\m$, $\Psi$ and
splitting \be A^\m\to A^\m+\ca^\m\ \ ,\  \psi \to \psi +\Psi \ ,\ee
we obtain
for the quadratic part of the action
 the expression of the type
\be S^{(2)}=\int d^4x {\pmatrix { \ca_\a & \bar \Psi_1 & \bar
\Psi_2 \cr} }\,\cb\, {\pmatrix{ \ca_\b \cr   \Psi_1 \cr  \Psi_2
\cr}} \ ,\ee where the matrix $\cb$ contains  classical fields. We
have to include in $\cb$ the gauge fixing term,  \be
S_{\rm{GF}}=-{1\over 2}\int d^4x(\pa_\m\ca^\m)^2\ ,\ee while the
ghost action will not contribute.
 The
one-loop effective action is  then
$$\G_1={i\over 2}\log\Sdet \cb = {i\over
2}\,\STr\log\cb \ .$$

Let us consider some properties of the matrix $\cb$. It can be
written in a 3$\times$3 block-matrix form
$$\cb =\pmatrix{ \cb _{11} &\cb _{12} &\cb _{13} \cr
\cb _{21} &\cb _{22} &\cb _{23} \cr \cb _{31} &\cb _{32} &\cb
_{33} \cr } \ ,$$ where the submatrices $\cb _{12}$, $\cb _{13}$,
$\cb _{21}$ and $\cb _{31}$ are Grassmann-odd while the rest are
Grassmann-even. The supertrace of $\cb$ is defined by
$$ \STr\cb = \Tr\cb _{11}-\Tr\cb _{22} -\Tr\cb _{33}\ .$$
$\cb$ is of the form
$$\cb =\pmatrix{ {1\over 2}g_{\a\b}\Box &0 &0 \cr
0 &i\slash \pa &0 \cr 0&0 &i\slash \pa \cr }+\cm \ .$$
 In order to expand $\log \cb$ around identity,
we multiply it by the matrix $\cc\cc^{-1}$ \cite{bv}, with
$$\cc =\pmatrix{ 2 & 0 & 0 \cr
0 & -i\slash\pa & 0  \cr 0 & 0 & -i\slash\pa \cr} \ .$$ Then
\bea\G_1&=&{i\over 2}\,\STr\log (\cb \cc)+{i\over 2}\,\STr\log
\cc^{-1}\ncr  &=&{i\over 2}\,\STr( \ci +\Box ^{-1}\cm\cc )+
{i\over 2}\,\STr\log \cc^{-1}+{i\over 2}\,\STr\log \Box  ,\ncr
\eea where $\ci ={\rm diag} (g_{\m\n},1,1)$. As usual, the second
and the third terms, being independent on the fields, can be
included in  infinite renormalization. Note that  the propagator
for all fields is now  $\Box ^{-1}$, while the massive fermionic
terms are in the interaction part, $\cm$.

Performing the transformations described above, for NC QED we
obtain the effective action in the following form: \be \Gamma
=S_0+{i\over 2}\,\STr\log \big(\ci +\Box ^{-1}N_0 +\Box ^{-1}N_1
+\Box ^{-1}T_1 +\Box ^{-1}T_2 \big) \ .\label{str} \ee The
matrices $N_0$, $N_1$, $T_1$ and $T_2$ are given by \be \label{n0}
N_0=\pmatrix {0& 0&0\cr 0&i m \slash\pa & 0\cr 0 &0 & i m\slash\pa
}\ ,\ee

\be\label{n1} N_1=\pmatrix {0& -i\bar\psi\g ^\a \slash\partial &
\bar\psi\g ^\a \slash\partial \cr 2\g ^\b\psi & 0 & \slash
A\slash\pa \cr -2i\g ^\b\psi& -\slash A\slash\pa & 0 }\ ,\ee

\be\label{t1} T_1=\pmatrix{V& A_1 & A_2\cr
 B_1 &  C & 0 \cr B_2 & 0 & C \cr }\ee

$$ T_2=\theta ^{\rho\sigma}\Delta^{\a\b\g}_{\sigma\r\m} \pmatrix
{-{1\over 2}\bar\psi\g ^\m\psi\pa _\g & {i\over 4}({1\over
2}F_{\g\b}+\pa _\g A_\b )\bar\psi\g ^\m \slash\partial &-{1\over
4}({1\over 2}F_{\g\b}+\pa _\g A_\b )\bar\psi\g ^\m \slash\partial
\cr - {1\over 2}\g ^\m\psi (-{1\over 2}F_{\g\a}+A_\a \pa _\g  )
&{i\over 8}A_\a F_{\g\b}\g^\m\slash\pa &-{1\over 8}A_\a
F_{\g\b}\g^\m\slash\pa  \cr {i\over 2}\g ^\m\psi (-{1\over
2}F_{\g\a}+A_\a \pa _\g  ) & {1\over 8}A_\a F_{\g\b}\g^\m\slash\pa
& {i\over 8}A_\a F_{\g\b}\g^\m \slash\pa \cr }$$
where $A_i,B_i\
(i=1,2)$ and $C$ are
 \bea A_i &=&{i\over 4}\theta
^{\r\s}\Delta ^{\a\b\g}_{\s\r\m}\pa _\g \Big(-(\pa _\b\bar\psi_i
)i\g ^\m-{m\over 2}\d ^\m _\b \bar\psi_i\Big)\slash\pa \ncr
 B_i
&=&{1\over 2}\theta ^{\r\s}\Delta ^{\a\b\g}_{\s\r\m} \Big( -i\g
^\m (\pa _\a\psi_i )+{m\over 2}\d ^\m _\a \psi_i\Big)\pa _\g \ncr
 C
&=&-{i\over 4}\theta ^{\r\s}\Big(- {i\over 2}\Delta
^{\a\b\g}_{\s\r\m}\g ^\m F_{\g\b}\pa _\a +m F_{\r\s}\Big)
\slash\pa \ .\nonumber\eea

 $V=
\overleftarrow { \pa _\m} V^{\m\a,\n\b}(x)\overrightarrow { \pa
_\b}$ \  comes from the term
$(\pa_\m\ca_\a)V^{\m\a,\n\b}(\pa_\n\ca_\b)$
in $S^{(2)}$ :
\bea \label{v}
V^{\m\r,\n\s}&=& {1\over 2}\,
(g^{\m\n}g^{\r\s}-g^{\m\s}g^{\n\r})\theta^{\a\b}F_{\a\b}\ncr
&+&g^{\m\n}(\theta^{\a\r}{F^\s}_\a+\theta^{\a\s}{F^\r}_\a)
+g^{\r\s}(\theta^{\a\m}{F^\n} _\a+\theta^{\a\n}{F^\m} _\a )\ncr
&-&g^{\m\s}(\theta^{\a\r}{F^\n}_\a+\theta^{\a\n}{F^\r}_\a)
-g^{\n\r}(\theta^{\a\s}{F^\m} _\a+\theta^{\a\m}{F^\s} _\a )\ncr
&+& \theta^{\m\r}F^{\n\s}+\theta^{\n\s}F^{\m\r}-
\theta^{\r\s}F^{\m\n} -\theta^{\m\n}F^{\r\s}
-\theta^{\n\r}F^{\m\s} -\theta^{\m\s}F^{\n\r} \ .\nonumber\eea We
will also use $V^{\m\n}_{\r\s}=g_{\r\a}g_{\s\b}V^{\m\a ,\n\b}$
which has the obvious symmetry $V_{\m\n}^{\a\b}=V_{\n\m}^{\b\a}$.

Note that in $N_1$ and $T_2$ the Majorana spinors $\psi _1$ and
$\psi _2$ add up neatly to the Dirac spinor $\psi$.
  $T_1$ depends on both of the Majorana spinors explicitly, but in
the same way. These properties ensure that the final results will
be expressed  in terms of the Dirac spinor.

\section{Divergent one-loop effective action}

The operator $\cb\cc$  in the formula (\ref{str}) is splitted in a
way convenient for the analysis of  perturbation series. Let us
explain the notation a little further.  $T$-matrices are linear in
the parameter $\theta$. Index
 denotes the number of classical fields in a given matrix,
i.e. in diagrammatic language, shows the number of 'external
legs' of the corresponding diagram. In our calculation we confine
to the corrections of linear and quadratic
 order in $\theta$ and of the second order in
classical fields. If we consider the expansion of (\ref{str}) \bea
\label{strexp} \G_1&=&{i\over 2}\,\STr\log \big( 1+\Box ^{-1}N_0
+\Box ^{-1}N_1 +\Box ^{-1}T_1 +\Box ^{-1}T_2 \big)\ncr &=&{i\over
2} \sum _{n=1}^\infty {(-1)^{n+1}\over n}\, \STr\big(\Box ^{-1}N_0
+\Box ^{-1}N_1 +\Box ^{-1}T_1 +\Box ^{-1}T_2 \big) ^n \ ,\eea it
seems as we have to include only $n=1,2$. But due to
the nonvanishing fermionic mass $m$ (i.e. the existence of
the term $N_0$), in principle we
will have to take into account also higher powers of $n$.
$n$ will be determined from the fact that we are
calculating only the divergent part. Analyzing the
structure of $\nn ,\dots ,\td $ in some detail we conclude that
the following terms in the expansion (\ref{strexp}) may be
divergent: $(\nn )^k\td  $ for $k=2,3,4$; $(\nn )^k\nj\tj$ for
$k=1,2,3$;
 $(\nn )^k(\nj )^2$ for $k=1,2$ and
 $(\nn )^k(\tj )^2$ for $k=1,2,3,4$.
(Here, of course, terms are written symbolically i. e. without the
exact order of the operators.)
 It is also clear that in the massless fermionic
 case the absence of $N_0$ brings a considerable simplification.

Let us first discuss maximally simplified situation: the purely
bosonic case. Then  $N_0$,
$N_1$ and $T_2$ are all absent, while $T_1$ reduces to $\ct _1$:
\be \ct _1 =\pmatrix{ V &0 &0\cr 0&0&0\cr 0&0&0\cr}\ .\ee  We have
then \be \G_b={i\over 2}\,\STr\log \big( 1+\Box ^{-1}\ct _1\big) =
{i\over 2}\Big[\Tr \Box ^{-1}\ct _1 -{1\over 2}\Tr (\Box ^{-1}\ct
_1 )^2+\dots\Big] \label{vv}\ee and in the $\theta ^2 ,A^2$-order we
need only first two terms. Even this calculation proves to be
technically difficult, mainly due to the complicated
tensorial structure of $V$.

 It can be easily seen that the divergent
part of the first trace in (\ref{vv}), $ \Tr\Box ^{-1}\ct _1$
vanishes. The second term in (\ref{vv}) is
\be\label{VV} \int
d^4x\, d^4y\, g^{\n\r}g^{\m\s}
V^{\a\b}_{\n\m}(x)V^{\g\d}_{\s\r}(y)\pa_\a^x\pa_\d^yG(y-x)\pa_\b^x\pa_\g^yG(x-y)\
,\ee where \be G(x-y)=-\int{d^4k\over (2\pi)^4}{e^{-ik(x-y)}\over
k^2} \ee is the Green function which satisfies \be
\Box_xG(x-y)=\d^{(4)}(x-y)\ .\ee
After  Fourier
transformation, the expression (\ref{VV}) becomes \be{1\over
(2\pi)^8}\int d^4p\, d^4k \, g^{\n\r}g^{\m\s}
V^{\a\b}_{\n\m}(p)V^{\g\d}_{\s\r}(-p){k_\b
k_\g(k+p)_\a(k+p)_\d\over k^2(k+p)^2}\ .\ee The dimensional
regularization in $D=4-\e$-dimensional space
gives for the divergent part of (\ref{VV})
 \bea &&{i\pi^{D\over
2}\over 2\e}\int{d^4p\over (2\pi)^8} g^{\n\r}g^{\m\s}
V^{\a\b}_{\n\m}(p)V^{\g\d}_{\s\r}(-p)\Big[{2\over 15} p_\a p_\b
p_\g p_\d\ncr &+&{1\over 15}p^2(g_{\a\b}p_\g p_\d +g_{\a\g}p_\b
p_\d+g_{\a\d}p_\b p_\g+g_{\g\b}p_\a p_\d+g_{\d\b}p_\a p_\g
+g_{\g\d}p_\a p_\d)\ncr &+&{1\over
60}(g_{\a\b}g_{\g\d}+g_{\a\d}g_{\b\g}+g_{\a\g}g_{\b\d})p^4\ncr
&-&{1\over 6}(g_{\b\g}p_\a p_\d -g_{\a\d}p_\b p_\g)p^2\Big]\
.\label{rezimp}\eea
Rewriting (\ref{rezimp}) in the coordinate space
we  obtain the one-loop
correction:
 \bea\G_{b}&=&{1\over
128\pi^2\e}\, g^{\n\r}g^{\m\s}\int d^4x\,\Big[{2\over
15}\,\pa_\a\pa_\b V^{\a\b}_{\n\m}\pa_\g\pa_\d
V^{\g\d}_{\s\r}-{1\over 3}g_{\b\g}\Box V^{\a\b}_{\n\m}\pa_\a\pa_\d
V^{\g\d}_{\s\r}\ncr &+&{2\over 15}\,(g_{\a\b}\Box
V^{\a\b}_{\n\m}\pa_\g\pa_\d V^{\g\d}_{\s\r} +g_{\a\g}\Box
V^{\a\b}_{\n\m}\pa_\b\pa_\d V^{\g\d}_{\s\r} +g_{\a\d}\Box
V^{\a\b}_{\n\m}\pa_\g\pa_\d V^{\g\d}_{\s\r})\ncr &+&{1\over 60}\,(
g_{\a\b}g_{\g\d}+g_{\a\d}g_{\b\g}+g_{\a\g}g_{\b\d})\Box
V^{\a\b}_{\n\m}\Box V^{\g\d}_{\r\s}\Big] \ .\eea The hard part now
 is to introduce the explicit form of
matrix elements (\ref{v}) and perform the index summations. The result
reads \bea\G_{b}\!\!&=&\!\!{1\over 64\pi ^2\e}\,\int d^4x \Big[
{12\over 5}\,\Box \tilde F_{\m\n}\Box \tilde F^{\m\n}+{2\over
5}\,\Box \tilde F_{\m\n}\Box \tilde F^{\n\m}-{1\over
60}\,(\Box\tilde F)^2\ncr\!\!& +&\!\!{7\over 15}\, \Box\tilde
F\pa_\m\pa_\n\tilde F^{\m\n} +{3\over 5}\,\Box \tilde
F_{\n\m}\pa^\n\pa_\s \tilde F^{\s\m} -{4\over 5}\,\Box \tilde
F_{\n\m}\pa^\n\pa_\s \tilde F^{\m\s}\ncr\!\!&
 -&\!\!{7\over 5}\,\Box \tilde
F_{\n\m}\pa^\m\pa_\s \tilde F^{\n\s}+{8\over
15}\,(\pa_\m\pa_\n\tilde F^{\m\n})^2 \ncr\!\!&-&\!\!{1\over
2}\,\theta^2\Box F^{\a\n}\pa_\a\pa_\d F^{\d}_{\ \n}-{1\over
2}\,\Box F_{\n\m}\tilde \pa^\a\tilde\pa_\a F^{\n\m}\Big]\
,\label{nesr}\eea
where we use the notation \be \tilde
F^{\m\n}={\theta ^\m}_\a F^{\a\n}\ ,\ \ \tilde F =\theta
_{\m\n}F^{\m\n}\ ,  \ \ \tilde\pa ^\m ={\theta ^\m}_\a \pa ^\a\
,\ \ \theta ^2=\theta ^{\r\s}\theta _ {\r\s}\ . \ee
 In order to simplify further
 we need the Bianchi identities; they are given in the
convenient form in the appendix. Finally, the divergent part of
the U(1)  effective action becomes\bea
\G_{b}\!\!&=&\!\!{1\over 64\pi^2\e}\,\int d^4x\Big[\Box \tilde
F^{\m\n}\Box\tilde F_{\m\n}+{2\over 15}\,\Box\tilde
F^{\m\r}\pa_\m\pa^\n\tilde F_{\n\r}\ncr\!\!&+&\!\!{1\over 5}\,\Box
F^{\m\n}\tilde \pa_a\tilde \pa^\a F_{\m\n}-{1\over 4}\,\theta
^2\Box F^{\m\n}\Box F_{\m\n}\Big]\ .\label{edkrajnje}\eea
This
result coincides exactly with the result of \cite{bggpsw}.
To conclude: working out the
bosonic part  separately we have not only
done a lengthy piece of calculation, but we have also performed an
important check.

Let us pass now to the full QED case.
 In the lowest nonvanishing order,
  corresponding
to $n=2$ in the expression (\ref{strexp}), for the correction
containing two classical fields we get
\bea \label{lprime}\G_1
^\prime &=&-{i\over 4}\, \STr\Big( \nn+\nj +\tj +\td\Big)^2\ncr
&=&-{i\over
4}\,\STr\Big((\nn +\nj )^2+2(\nn +\nj )\tj+(\tj)^2\Big) \ .\eea
The relevant supertraces (i.e., their ${1\over\e}$-parts) are:
 \be \label{11}\STr(\nn +\nj )^2 ={i\over (4\pi
)^2\epsilon}\Big( -{8\over 3}F_{\m\n}F^{\m\n}+16
i\bar\psi\slash\pa\psi \Big) \ ,\ee
\be 2\STr (\nn +\nj )\tj
=\cpe\theta ^{\rho\s} \Big[ {4\over 3}\bar\psi\g _\r\pa _\s (\Box
-im\slash\pa )\psi +{2\over 3}\bar\psi\s _{\r\s}\Box (i\slash\pa
-m)\psi\Big] \ ,\ee \bea \STr(\tj)^2 &=&\cpe\Big[{i\over 12}\Big(
2{\theta ^\r}_\m\theta ^{\m\s}\bar\psi\pa _\r\pa _\s \slash\pa
\Box\psi +4{\theta ^\r}_\m\theta ^{\m\s} \bar\psi\g _\r\pa _\s\Box
^2\psi + \theta ^2\bar\psi\slash\pa\Box ^2\psi \ncr &+&2im\theta
^2\bar\psi\Box ^2\psi \ncr &-&2m^2{\theta ^\r}_\m\theta ^{\m\s}
\bar\psi\pa _\r\pa _\s\slash\pa\psi -2m^2{\theta ^\r}_\m\theta
^{\m\s} \bar\psi\g _\r\pa _\s\Box\psi-m^2\theta
^2\bar\psi\slash\pa\Box\psi \Big)\ncr &-& {m^2\over 2}\,\tilde
F\Box\tilde F -{1\over 30}\,\tilde F^{\r\s}\Box ^2\tilde
F_{\s\r}+{2\over 15}\,\tilde F^{\r\s}\Box ^2\tilde F_{\r\s}
-{2\over 15}\,\tilde F^{\r\s}\Box\pa _\m\pa^\n\tilde F_{\r\n}
\Big] \ncr &+& VV {\rm term}\ . \label{22}\eea
$VV$ is the bosonic
correction previously found. $\Gamma _1^\prime$ is now easily
obtained summing up (\ref{11}-\ref{22}).

The list of
supertraces necessary to find the higher contributions, $n=3,\dots$,
 is:
\be \STr (\nn)^k\td =0\  \ {\rm for}\ k=1,2,3,4\ ,\ee \be
\STr\nn(\nj)^2= \cpe 32m\int d^4x\bar\psi\psi \ ,\ee \be
4\STr(\nn)^2(\nj)^2+2\STr\nn\nj\nn\nj =0\ ,\ee
$$\STr\Big[(\nn)^2\nj\tj+(\nn)^2\tj\nj+\nn\nj\nn\tj\Big]$$
\be =\cpe
\theta ^{\r\s}\int d^4 x\Big( 2m^2\bar\psi\g
 _\r\pa _\s\psi +im^2\bar\psi\g _\r\g _\s (i\slash\pa
 -m)\psi\Big)\ ,\ee
\bea\STr\nn(\tj)^2&=&\cpe\int d^4 x\Big({m^3\over 3}{\theta
^\r}_\m\theta ^{\m\s}
 \bar\psi\pa _\r\pa _\s\psi+{m\over 6}{\theta ^\r}_\m\theta
^{\m\s}
 \bar\psi\pa _\r\pa _\s\Box\psi\\ \nonumber
&-&{im^2\over 3}{\theta ^\r}_\m\theta
^{\m\s}
 \bar\psi\gamma _\r\pa _\s\Box\psi+
{m^3\over 12}\,\theta ^2\bar\psi\Box \psi-{im^2\over 6}\,\theta
^2\bar\psi\Box \slash\pa\psi\\ \nonumber
&-&{m\over 12}\,\theta
^2\bar\psi\Box ^2\psi -{m^2\over 3}\tilde F\Box\tilde F\Big)\
,\eea
\bea\STr(\nn)^2(\tj)^2&=&\cpe m^2\int d^4x \Big({m\over 3}
{\theta ^\r}_\m\theta ^{\m\s}
 \bar\psi\pa _\r\pa _\s\psi
-{m\over 6}\theta ^2\bar\psi\Box \psi \\ \nonumber
&-&{im^2\over
3}\,\theta^{\r}_{\ \m}\theta^{\m\s}\bar\psi\g_\s\pa_\r\psi +{i\over
6}\,{\theta ^\r}_\m\theta ^{\m\s}
 \bar\psi\pa _\r\pa _\s\slash\pa\psi\\ \nonumber
&+&{i\over 3}\,
 {\theta ^\r}_\m\theta
^{\m\s}
 \bar\psi\g _\r\pa _\s\Box\psi+{i\over 12}\,
\theta ^2\bar\psi\slash\pa\Box\psi -{im^2\over
12}\theta ^2\bar\psi\slash\pa\psi \ncr
&-&m^2\tilde F^2-{1\over 12}\,
\tilde
F\Box\tilde F +{1\over 2}\,\tilde F^{\r\s} \Box\tilde F_{\r\s}\\
\nonumber &-&{1\over 3}\,\tilde F^{\r\s}\pa _\s\pa ^\n\tilde
F_{\r\n} -{1\over 6}\tilde F^{\r\s} \Box\tilde F_{\s\r}\Big) \
,\eea
\be \STr\nn\tj\nn\tj =\cpe\int d^4 x\Big(-{m^2\over 3}
\tilde F^{\r\s} \Box\tilde F_{\r\s}-m^4\tilde F^2\Big)\ ,\ee
\bea\STr(\nn)^3\tj^2&=&-\cpe{m^3\over 4}\int d^4x\Big( -2{\theta
^\r}_\m\theta ^{\m\s}
 \bar\psi\pa _\r\pa _\s\psi\\ \nonumber
&+&4im{\theta
^\r}_\m\theta ^{\m\s}
 \bar\psi\g _\r\pa _\s\psi-m^2\theta^2\bar\psi\psi
+\theta^2\bar\psi\Box\psi\\ \nonumber
&+&2im\theta^2\bar\psi\slash\pa\psi+4m\tilde F^2\Big)\ ,\eea
\be\STr(\nn)^2\tj\nn\tj =-\cpe m^4\int d^4x \tilde F^2\ ,\ee
\bea\STr(\nn)^4(\tj)^2&=&\STr((\nn)^2\tj(\nn)^2\tj)\\
\nonumber &=&-\cpe m^4\int d^4x\Big(-{2\over 3}\tilde
F_{\m\n}\tilde F^{\m\n}+{1\over 3}\tilde F^2+{1\over 3} \tilde
F_{\m\n}\tilde F^{\n\m}\Big)\ ,\eea \be\STr(\nn)^3(\tj)\nn\tj
=-\cpe m^4\int d^4x\tilde F_{\m\n}\tilde F^{\m\n}\ ,\ee

Expanding the logarithm under the supertrace and
using the previous results for the higher-order contribution
we obtain:
 \bea\G_1^{\prime\prime}&=&\cpe{1\over 2}\int d^4
 x\Big[ 32mi\bar\psi\psi-im^2
\theta ^{\r\s}( 2\bar\psi\g
 _\r\pa _\s\psi
+i\bar\psi\g _\r\g _\s (i\slash\pa
 -m)\psi)\ncr
 &+&{im^3\over 2}
{\theta ^\r}_\m\theta ^{\m\s}
 \bar\psi\pa _\r\pa _\s\psi+{2m^2\over 3}\theta ^\r _{\ \m}\theta^{\m\s}
 \bar\psi\gamma _\r\pa _s\Box\psi\ncr
&+&{m^2\over
 4}\theta^2\bar\psi\Box\slash\pa\psi-{im\over
 12}\theta^2\bar\psi\Box^2\psi+{5m^4\over
 12}\theta^2\bar\psi\slash\pa\psi\ncr &+&{im\over 6}{\theta ^\r}_\m\theta
^{\m\s}
 \bar\psi\pa _\r\pa _\s\Box\psi+{2m^4\over 3}{\theta ^\r}_\m\theta
^{\m\s}
 \bar\psi\g _\r\pa _\s\psi+{m^2\over 6}{\theta ^\r}_\m\theta
^{\m\s}
 \bar\psi\pa _\r\pa _\s\slash\pa\psi\ncr
 &+&{im^5\over 4}\theta^2\bar\psi\psi -{im^2\over 4}\, \tilde
F\Box\tilde F -{im^2\over 3}\,\tilde F^{\r\s} \Box\tilde
F_{\r\s}\ncr &+&{im^2\over 3}\,\tilde F^{\r\s}\pa _\s\pa
^\n\tilde F_{\r\n} +{im^2\over 6}\tilde F^{\r\s}\Box \tilde
F_{\s\r}+{im^4\over 2}\tilde F_{\r\s}\tilde F^{\s\r}\Big] \
,\label{lsec}\eea

Adding (\ref{lprime}) and (\ref{lsec}) we get
the divergent part of the full one-loop corrected effective action
  for NC QED:

\bea\G_1&=&{1\over (4\pi)^2\e}\int d^4 x
\Big[ 4i\bar\psi\slash\pa\psi-16m\bar\psi\psi-{2\over
3}F_{\m\n}F^{\m\n}\\ \nonumber
&+& \theta^{\a\b}\Big( {1\over
3}\,\bar\psi\g _\a\pa _\b (\Box -im\slash\pa )\psi +{1\over
6}\bar\psi\s _{\a\b}\Box (i\slash\pa -m)\psi\\ \nonumber
&+&m^2\bar\psi\g_\a\pa_\b\psi+{m^2\over 2}\,\bar
\psi\s_{\a\b}(i\slash\pa-m)\psi \Big)\\ \nonumber
&-&{1\over 120}\,\tilde F^{\r\s}\Box ^2\tilde F_{\s\r}+{1\over 30}\,\tilde
F^{\r\s}\Box ^2\tilde F_{\r\s} -{1\over 30}\,\tilde
F^{\r\s}\Box\pa _\s\pa^\n\tilde F_{\r\n}\\ \nonumber
 &+&{m^2\over
6}\,\tilde F^{\r\s}\Box \tilde F_{\r\s}-{m^2\over 12}\,\tilde
F^{\r\s}\Box \tilde F_{\s\r} -{m^2\over 6}\,\tilde
F^{\r\m}\pa _\m\pa^\n\tilde F_{\r\n} -{m^4\over 4}\,\tilde
F_{\r\s}\tilde F^{\s\r}\\ \nonumber &+&{i\over
48}\,\theta^2\bar\psi\Box^2\slash\pa\psi-{i\over
24}\,\theta^{\a\m}\theta^\b_{\
\m}\bar\psi\Box\slash\pa\pa_\a\pa_\b\psi -{i\over
12}\,\theta^{\a\m}\theta^\b_{\ \m}\bar\psi\Box^2\g_\a\pa_\b\psi \\
\nonumber &+&{m\over 12}\,\theta^{\a\m}\theta ^b_{\
\m}\bar\psi\pa_\r\pa_\s\Box\psi \\ \nonumber &+&{5im^2\over
48}\,\theta^2\bar\psi\Box\slash\pa\psi-{im^2\over 24}\,
\theta^{\a\m}\theta^\b_{\ \m}\bar\psi\slash\pa\pa_\a\pa_\b\psi
-{7im^2\over 24}\,\theta^{\a\m}\theta^\b_{\
\m}\bar\psi\g_\a\pa_\b\Box\psi\\
\nonumber
&+&{m^3\over
4}\,\theta^{\a\m}\theta^\b_{\ \m}\bar\psi\pa_\a\pa_\b\psi\\
\nonumber
&+&{5im^4\over
24}\,\theta^2\bar\psi\slash\pa\psi-{im^4\over
3}\theta^{\a\m}\theta^\b_{\
\m}\bar\psi\g_\a\pa_\b\psi \\ \nonumber
&-&{m^5\over 8}\,\theta ^2\bar\psi\psi
\Big]+\G_{b}\ .\label{krajnje}\eea

\section{Conclusions and outlook}

Our goal in this paper was to obtain the
divergent part of the one-loop effective action  in NC QED in the
second order in the noncommutativity parameter $\theta$ and the same
order in the classical fields, $\psi, A^\m$. Thus we obtained the
second order corrections
to the propagators in the theory and therefore the form of
counterterms necessary for renormalization. The method
we used is the
background field method; the initial point for the perturbative
expansion is (\ref{str}). It is written in such a way  that it
is easy to sample out the terms contributing to the 2-point,
3-point, 4-point etc. functions. For example, the terms with 3
external fields (corrections to 3-vertices)
 are, in the linear order in $\theta$: $(\nn )^k(\nj )^2\tj$ and
$(\nn )^k\nj \td$;
in the quadratic order in $\theta$: $(\nn )^k\nj (\tj )^2$.
 $k$ can be determined from the condition that the corresponding
 integrals are divergent.
 The corrections to the 4-point functions are given by the
 terms: $\theta$-linear:
 $(\nn )^k(\nj )^3\tj$ and  $(\nn )^k(\nj )^2\td$;    \
 $\theta$-quadratic: $(\nn )^k(\nj )^2(\tj )^2$ and  $(\nn )^k(\td )^2$.

Because of the length of calculations we confined ourselves in this
paper to the corrections to the 2-point functions. However, in some cases it is not
difficult to extract other results. E.g., the divergent part of the
4-fermion vertex, in the linear order in $\theta$, is easily found to be
\be\label{4psi} S_{4\psi} =-\jcpe {1\over 2}\theta ^{\r\s}\DG\int d^4x\, \bar\psi\g ^\m\psi\,
\bar\psi \g _ \a\g _\b\g _\g\psi \ ,\ee
in agreement with \cite{bggpsw}  .

As we have already pointed out, our main motive was to check
 the renormalizability of $\theta$-expanded NC QED in the first
and second order in $\theta$ and the possibilities of
 generalization  to all orders. This
was done for the pure NC U(1) in \cite{bggpsw}. The trick which was used is
that the SW map does not fix the fields
in the $\theta$-expansion fully, but allows for their redefinitions.
If the fields are expanded as (symbolically written)
\be \label{SWexpansion}\hat A_\m=\sum \theta ^n A_\m^{(n)}\ ,\ \
\hat \psi =\sum \theta ^n \psi ^{(n)}\ ,\ee
the allowed redefinitions are of the form
\be\label{reda} {A_\m^{(n)}}^\prime =A_\m^{(n)} + {\bf A}_\m^{(n)}\ ,\ee
\be\label{redpsi} {\psi_\m^{(n)}}^\prime =\psi _\m^{(n)} + {\bf \Psi}_\m^{(n)}\ ,\ee
where ${\bf A}_\m^{(n)}$, ${\bf \Psi}_\m^{(n)}$
are gauge covariant expressions of appropriate dimension with exactly
$n$ factors of $\theta$. These field redefinitions produce in the action
 extra terms of the following forms \cite{bggpsw}:
\be \label{DSA} \Delta S_A =\int d^4x\, (D_\n F^{\m\n}){\bf A}_\n^{(n)}\ee
\be \label{DSpsi} \Delta S_\psi =\int d^4x\,
\Big[ \bar\psi (i\slash D -m){\bf \Psi}^{(n)}+
\bar {\bf \Psi}^{(n)}(i\slash D -m)\psi \Big]\ .\ee
So if the renormalizability of the theory can be achieved by the field
redefinitions, all counterterms have to be of the types (\ref{DSA}-\ref{DSpsi}).

First, it is easy to see that in the purely bosonic case the
action (\ref{edkrajnje}) is of the type (\ref{DSA}). Let us
discuss what happens when the fermions are present. All bosonic
corrections are of the $\theta ^2$-order, and all are of the
allowed type except for the term $-\jcpe {m^4\over 4}\tilde
F_{\r\s}\tilde F^{\s\r}$ . The fermionic $\theta$-linear
correction is \bea  & & \jcpe \theta^{\a\b}\Big[
{1\over 3}\,\bar\psi\g _\a\pa _\b (\Box -im\slash\pa )\psi
+{1\over 6}\bar\psi\s _{\a\b}\Box (i\slash\pa -m)\psi\\ \nonumber
&+&m^2\bar\psi\g_\a\pa_\b\psi+{m^2\over 2}\,\bar
\psi\s_{\a\b}(i\slash\pa-m)\psi \Big]\ .\eea One can check that
the pieces \ $ {1\over 3}\,\bar\psi\g _\a\pa _\b \Box \psi
+{1\over 6}\,\bar\psi\s _{\a\b}\Box (i\slash\pa -m)\psi$\  and \
$m^2\bar\psi\g_\a\pa_\b\psi+{m^2\over 2}\,\bar
\psi\s_{\a\b}(i\slash\pa-m)\psi $\  are allowed by field
redefinitions, leaving the term\  $-{im\over 3}\bar\psi\gamma
_\r\pa _\s\psi$ \ excessive. Thus we see that, indeed,
renormalizability cannot be achieved in the massive case, $m\neq
0$, as there are two terms (one bosonic and one fermionic) which
prevent it. Both terms are proportional to $m$.

One observes further that for $m=0$ also the $\theta ^2$
fermionic contribution fits into the field redefinition scheme.
Namely this contribution reads \be \jcpe\Big[ {i\over
48}\,\theta^2\bar\psi\Box^2\slash\pa\psi-{i\over
24}\,\theta^{\a\m}\theta^\b_{\
\m}\bar\psi\Box\slash\pa\pa_\a\pa_\b\psi -{i\over
12}\,\theta^{\a\m}\theta^\b_{\ \m}\bar\psi\Box^2\g_\a\pa_\b\psi
\Big]\ ,\ee and it is of the form (\ref{DSpsi}) for ${\bf
\Psi}^{(2)}$ given by
\be {\bf \Psi}^{(2)}= \jcpe\Big[ {1\over 96}\,\theta^2 \Box^2\psi-{1\over
48}\,\theta^{\a\m}\theta^\b_{\ \m}\Box\pa_\a\pa_\b\psi -{1\over
24}\,\theta^{\a\m}\theta^\b_{\ \m}\Box\slash\pa\g_\a\pa_\b\psi
\Big]\ .\ee

For completeness, let us rewrite the effective action in the
massless case: \bea\G_{1,m=0}&=&{1\over (4\pi)^2\e}\int d^4 x
\Big[ 4i\bar\psi\slash\pa\psi-{2\over 3}F_{\m\n}F^{\m\n}\\
\nonumber &+& \theta^{\a\b}\Big( {1\over 3}\,\bar\psi\g _\a\pa _\b
\Box\psi +{1\over 6}\,\bar\psi\s _{\a\b}\Box i\slash\pa \psi\Big)
\\ \nonumber &-&{1\over 120}\,\tilde F^{\r\s}\Box ^2\tilde
F_{\s\r}+{17\over 60}\,\tilde F^{\r\s}\Box ^2\tilde F_{\r\s}+
{1\over 20}\, F^{\r\s}\Box\tilde\pa _\m\tilde\pa^\m F_{\r\s}
-{1\over 16}\theta ^2 F^{\r\s}\Box ^2F_{\r\s}\\ \nonumber
&+&{i\over 48}\,\theta^2\bar\psi\Box^2\slash\pa\psi-{i\over
24}\,\theta^{\a\m}\theta^\b_{\
\m}\bar\psi\Box\slash\pa\pa_\a\pa_\b\psi -{i\over
12}\,\theta^{\a\m}\theta^\b_{\ \m}\bar\psi\Box^2\g_\a\pa_\b\psi
\Big]\ .\eea

To summarize: our calculation inforces the conclusions
of \cite{w} (to the $\theta^2$-order): the propagators in
NC QED can be renormalized by the field redefinitions if
 fermions are massless; for massive fermions this is not possible.
Unfortunately, as also stressed in \cite{w}, the full
renormalizability is spoiled by the term
 (\ref{4psi}) arousing in the correction of 4-fermions vertex.
This correction
is not of the form (\ref{DSpsi}) (has no derivatives),
and therefore cannot be obtained
by field redefinitions.

One may expect that the NC versions of nonabelian gauge theories
have different properties from NC U(1). On the other hand,
the similar approach to the problem of renormalizability
seems to be viable, at least for $SU(2)$ Yang-Mills theory.
This will be the subject of the fortcoming publications.

\section{Appendix}

Using Bianchi identities we obtained the following expressions:

\be \pa_\r\pa_\s \tilde F^{\r\s}=-{1\over 2}\Box \tilde F\ ,\label{id2}\ee
 \be  \tilde \pa^\a F^{\r\s}\tilde\pa_\a
 F_{\r\s}=-2\tilde F^{\m\n}\Box \tilde F_{\m\n}+2 \tilde
 F^{\a\s}\pa_\rho \pa _\s\tilde F_\a^{\ \r} \label{id4}\ee
 \be\tilde F^{\m\r}\pa^\n\pa_\r\tilde F_{\n\m}=\tilde
F^{\m\r}\pa_\m\pa^\n\tilde F_{\n\r}=-{1\over 2}\tilde F
\pa_\m\pa_\n \tilde F^{\m\n}\ ,\label{id5}\ee
\be \tilde F^{\m\r}\pa^\n\pa_\m\tilde F_{\r\n}={1\over 2}\tilde F_{\m\n}\Box
\tilde F^{\n\m}\ ,\ee
\be F^{\a\n}\pa_\a\pa^\m F_{\m\n}={1\over
2}F^{\m\n}\Box F_{\m\n} \label{id6}\ .\ee

Identities including $\DG$:

\be \DG \d ^\m _\a =2(\d ^\b _\s \d ^\g _\r -\d ^\g _\s \d ^\b _\r   )\ee
\be \theta ^{\r\s}\theta^{rs}\DG\DR \d ^\m _\a\d ^m_a = 16\theta ^{\g\b}\theta ^{cb}\ee
 \be \theta ^{\r\s}\theta^{rs}\DG\DR \d ^m _\a\d ^\m_a = 8(\theta ^{b\b}\theta ^{c\g}
 - \theta ^{c\b}\theta ^{b\g})\ee
  \be \theta ^{\r\s}\theta^{rs}\DG\DR  g_{b\b}g_{c\g}g^{m\m}=
  -8{\theta ^\a}_\r \theta ^{\r\a}+4\theta ^2g^{a\a} \ee

\end{document}